\documentstyle[aps,prl,epsfig,floats]{revtex}

\begin{document}
\draft
\wideabs{

\title {Field-induced spin-density-wave phases in TMTSF organic
conductors: quantization versus non-quantization} 

\author{K. Sengupta$^{(1)}$ and N. Dupuis$^{(2)}$  }

\address{(1) Department of Physics, Yale university, New Haven,
CT-06520-8120 \\
(2) Laboratoire de Physique des Solides, CNRS UMR 8502,
  Universit\'e Paris-Sud, 91405 Orsay, France }

\date{February 7, 2003}

\maketitle

\begin{abstract}
We study the magnetic-field-induced spin-density-wave (FISDW) phases
in TMTSF organic conductors in the framework of the quantized nesting
model. In agreement with recent suggestions, we find that the SDW
wave-vector ${\bf Q}$  deviates from its quantized value near the
transition temperature $T_c$  for all phases with quantum numbers
$N>0$. Deviations from 
quantization are more pronounced at low pressure and higher $N$ and
may lead to a suppression of the first-order transitions $N+1\to N$
for $N\ge 5$.  Below a critical
pressure, we find that the $N=0$ phase invades the entire phase diagram
in accordance with earlier experiments. We also show that at $T=0$,
the quantization of ${\bf Q}$ and hence the Hall conductance is
always exact. Our results suggest a novel phase transition/crossover at
intermediate temperatures between phases with quantized and
non-quantized ${\bf Q}$. 

\end{abstract}

 }
%%%%%%%%%%%%%%%%%%%%%%%%%%%%%%%%%%%%%%%%%%%%%%%%%%%%%%%%%%%%%%%%%%%%%%

{\it Introduction.} 
Quasi-one-dimensional (Q1D) organic conductors of the $\rm (TMTSF)_2X$
family \cite{note1} (also known as the Bechgaard salts) are highly
anisotropic crystals that consist of parallel conducting chains. The
electron transfer integrals along the chains (in the $a$ direction)
and transverse to the chains (in the $b$ and $c$ directions) are $t_a
= 250$ meV, $t_b = 25$ meV, and $t_c = 0.75$ meV \cite{Ishi1}. Because
of the strong anisotropy, the Fermi surface of these materials
is open and consists of two disconnected sheets located near $\pm
k_F$, which are the Fermi momenta along the chains. In the presence of
a moderate magnetic field $H$ along the $c$ axis, the interplay between
the nesting property of the open Fermi surface and the quantization of
electron orbits due to the magnetic field leads to a cascade of
magnetic-field-induced spin-density-wave (FISDW) phases \cite{Ishi1}.
These phases have long been theoretically explained in the framework
of the quantized nesting model (QNM)
\cite{Ishi1,Gorkov84,Heritier84,Yamaji85,Montambaux85,Virosztek86,Mon1}.
A central prediction of the QNM
is that within each FISDW phase characterized by an integer $N$, the
wave vector ${\bf Q}=(Q_x,Q_y)$ of the spin modulation is
quantized: $Q_x =2k_F+NG$, where $G= ebH/\hbar c$ is the magnetic
wave vector and $b$ the interchain distance. As the field increases,
the integer $N$ varies, which leads to the FISDW cascade 
($N=\cdots,4,3,2,1,0$). In each phase, the quantization of
$Q_x$ implies the quantization of the Hall effect:
$\sigma_{xy}=-2Ne^2/h$ per layer of TMTSF molecules
\cite{Poilblanc87,Yakovenko91}. The ability of the QNM to explain the
quantum Hall effect (QHE) observed in the Bechgaard salts
\cite{Cooper} is one of its main successes. 

Recently Lebed' called into question some fundamental aspects of
the QNM \cite{Lebed}. He showed that due to the particle-hole
asymmetry in the FISDW phases with $N\ne0$, $Q_x$ deviates from its
quantized values. At the metal-FISDW transition, deviations from
quantization are controlled by the ratio $h=\omega_c/\pi T_c$ where
$\omega_c=v_FG$ ($v_F$ is the Fermi velocity along the chains) and
$T_c$ is the transition temperature. When $h$ reaches a critical value
$h_c$, the first-order transitions between different FISDW phases are
suppressed. $Q_x$ then becomes a continuous function of the
field. At
lower temperatures, first-order transitions (i.e. discontinuous jumps
of $Q_x$) survive although $Q_x$ is not quantized. Lebed's results
call into question our theoretical 
understanding of the QHE in the Bechgaard salts, since the latter
relies on the quantization of the  FISDW wave vector
\cite{Poilblanc87,Yakovenko91,note2}. 

Lebed's conclusions raise two important questions. First, the very
existence of the FISDW phases, which is due to a {\it quantum} effect
of the field, requires $h=\omega_c/\pi T_c$ to be large enough.
Indeed, when $T\gg \omega_c$, the magnetic field can be
treated semiclassically and we expect the FISDW cascade to disappear
in favor of either the metallic phase or the SDW with $Q_x=2k_F$
(i.e. the phase $N=0$). Thus, we expect the suppression of first-order
phase transitions to occur only in a small window of the 
parameter $h$. Second, the fate of the QHE can be understood only by
considering explicitly the low-temperature limit. The extrapolation of results
valid near $T_c$, as done by Lebed', is not reliable since the SDW
wave vector $Q_x$ may vary with temperature. 

In this Letter, we investigate the FISDW phase diagram both at $T=T_c$
and $T=0$ as a function of the strength of the electron-electron
interaction. The latter is a decreasing function of pressure and can
therefore be varied experimentally. We find that $Q_x$
deviates from its quantized value near $T_c$ for all phases $N>0$.
Deviations from quantization are stronger at low pressure and higher
$N$. When pressure is decreased, suppression of first-order phase
transitions occurs for $N\geq 5$. At lower pressure, below a critical
value $P_c$,  we find that
the $N=0$ phase invades the entire phase diagram in accordance with
earlier experiments \cite{Kang1}. On the other hand, at $T=0$ the 
quantization of $Q_x$ and hence the Hall conductance is exact for
all pressures and all $N$, down to the critical pressure $P_c$ below which
the $N=0$ phase again invades the phase diagram. Our results suggest a novel
phase transition/crossover at intermediate temperature
between phases with quantized and non-quantized $Q_x$. 

{\it Metal-FISDW transition.} 
The Hamiltonian
describing the Bechgaard salts in the vicinity of the Fermi energy in
the presence of a magnetic field ${\bf H}= H\hat z$ can be written as
\begin{eqnarray}
{\mathcal H}&=& \sum_{\alpha,\sigma} 
\int d^2r \psi^{\dagger}_{\alpha
\sigma}({\bf r}) [ v_F( -i \alpha \partial_x - k_F) 
\nonumber \\ &&  
+ t_\perp(-ib \partial_y-Gx)] \psi_{\alpha \sigma} ({\bf r}) 
\nonumber \\
&& + \frac{g}{2} \sum_{\alpha,\sigma,\sigma'} 
\int d^2r \psi^{\dagger}_{\alpha
\sigma}({\bf r})\psi^{\dagger}_{{\bar \alpha}
\sigma'}({\bf r}) \psi_{{\bar \alpha}
\sigma'}({\bf r})\psi_{\alpha
\sigma}({\bf r}). 
\label{interh}
\end{eqnarray} 
Here the operator $\psi_{\alpha\sigma}^{(\dagger)}({\bf r})$ creates
(annihilates) a right ($\alpha=+$) or left ($\alpha=-$) moving electron
with spin $\sigma$. We use the notation ${\bf r}=(x,mb)$ ($m$ integer)
and $\int d^2r=b\sum_m \int dx$. $v_F =\sqrt 2 t_a a $ is the Fermi
velocity along the chains (with $t_a$ the hopping amplitude and $a$
the lattice spacing) and
$g$ the amplitude of the electron-electron interaction. We have
linearized the Hamiltonian around the Fermi energy and used the gauge
${\bf A}=(0,Hx,0)$. $t_{\perp}(u)= -2t_b \cos(u)-2t_{2b} \cos(2u)$ 
describes the interchain hopping in a tight-binding
approximation, $t_b$ being the nearest-neighbor hopping. 
The next-nearest neighbor hopping amplitude $t_{2b}$
destroys the perfect nesting of the Fermi
surface and stabilizes the metallic phase in the absence of magnetic
field. Here and in the rest of this work $\hbar=c=1$.
%%%%%%%%%%%%%%%%%%%%%%%%%%%%%%%%%%%%%%%%%%%%%%%%%%%%%%%%%%%
\begin{figure}
\centerline{\psfig{file=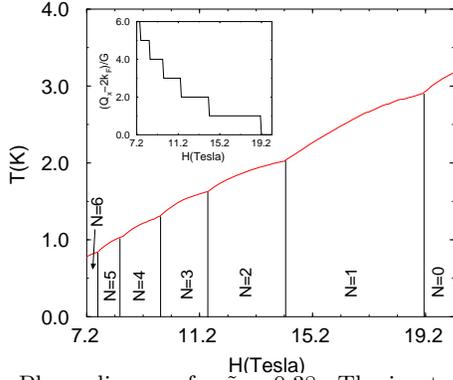,width=0.7\linewidth,angle=0}}
\caption{Phase diagram for $\tilde g=0.38$. The inset shows the
  parameter $N=(Q_x-2k_F)/G$ at the metal-FISDW transition ($T=T_c$) as a
  function of the magnetic field. The vertical 
lines are guides to the eyes and indicate first-order transitions.}
\label{fig1}
\end{figure}
\noindent
%%%%%%%%%%%%%%%%%%%%%%%%%%%%%%%%%%%%%%%%%%%%%%%%%%%%%%%%%%%
%%%%%%%%%%%%%%%%%%%%%%%%%%%%%%%%%%%%%%%%%%%%%%%%%%%%%%%%%%%
\begin{figure}
\centerline{\psfig{file=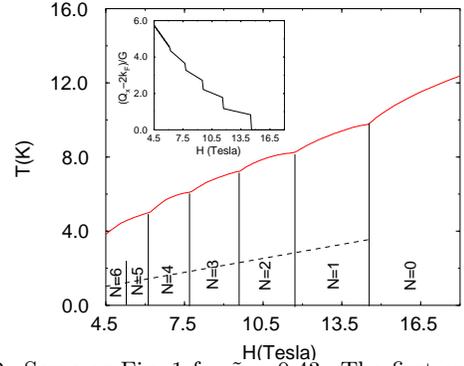,width=0.7\linewidth,angle=0}}
\caption{Same as Fig.~\ref{fig1} for $\tilde g=0.43$.
%The circles in the inset are analytical estimate of $Q_x$. 
The first-order transition line between the phases $N=6$ and $N=5$
terminates by a second-order critical point above which $N$ varies
continuously with the field. The dashed line schematically indicates a
phase transition/crossover at intermediate temperature between phases
with quantized and non-quantized $Q_x$ (see text).  
}
\label{fig2}
\end{figure}
\noindent
%%%%%%%%%%%%%%%%%%%%%%%%%%%%%%%%%%%%%%%%%%%%%%%%%%%%%%%%%%%
To obtain the phase diagram near $T_c$, we compute the static
spin susceptibility $\chi({\bf q})$ within the random-phase
approximation:
$\chi ({\bf q}) = \chi_0 ({\bf q})/(1-g \chi_0({\bf q}))$ 
where $\chi_0({\bf q})$ is the bare spin susceptibility. It can be
written as \cite{Mon1}
\begin{eqnarray}
\chi_0({\bf q}) &=& \sum_{n=-\infty}^{\infty} 
I^2_n(q_y) \chi_{\rm 1D}(q_x - n G) ,
\label{static} \\
\chi_{\rm
1D}(q_x) &=& \frac{N(0)}{2} \Bigg[ \ln \left(\frac{2\gamma
E_0}{\pi T}\right) + \Psi\left(\frac{1}{2}\right) \nonumber\\
&& -{\rm Re} \Psi\left(\frac{1}{2} + \frac{v_F}{4i\pi T}(q_x -
2k_F)\right) \Bigg] ,
\label{static1D} \\
I_n(q_y)&=&\int^{2\pi}_0 \frac{du}{2\pi} e^{inu +
\frac{i}{\omega_c}[T_{\perp}(u+q_yb/2)
+ T_{\perp}(u-q_yb/2)]}   , 
\label{form} 
\end{eqnarray}
where $T_{\perp}(u)=\int_0^{u} du' t_{\perp}(u')$, $N(0)= 1/\pi v_F b$
is the density of states per spin, $\Psi$ the digamma function,
$E_0$ an ultraviolet cutoff of the order of the bandwidth, and
$\gamma\simeq 1.781$ the exponential of the Euler constant. The
instability to the FISDW phase occurs when the Stoner criterion
$1-g\chi_0({\bf q})=0$ is satisfied. 

Since the 1D susceptibility has a logarithmic divergence for
$q_x=2k_F$, Eq.~(\ref{static}) suggests that the SDW instability
will occur with a quantized wave vector ${\bf Q}=(2k_F+NG,Q_y)$ ($N$
integer). However, the parameter 
$N=(Q_x-2k_F)/G$ obtained from the Stoner criterion is not in general
an integer. This can easily be shown analytically. Writing
$Q_x=2k_F+(N+\epsilon)G$ with $N$ integer and $\epsilon\ll 1$, the maximum
of the susceptibility $\chi_0$ (which will give the highest transition
temperature) is found to be determined by
\begin{eqnarray}
\epsilon(T,Q_y)= \frac{8\pi^2 T^2}{7 \zeta(3)\omega_c^2} \sum_{n \ne 0}
\frac{I_{N+n}^2}{n I_N^2},
\label{epsilon}
\end{eqnarray}
where $\zeta$ is the Riemann Zeta function and $I_n\equiv I_n(Q_y)$.
$\epsilon$ vanishes only
in the phase $N=0$ (for which $Q_y=\pi/b$) due to the particle-hole
symmetry which implies $I_n(\pi/b)=I_{-n}(\pi/b)$. 
%We find that the analytical formula matches 
%the numerical result quite well for $N\le 5$ when $\epsilon/N \ll
%1$, but deviates from it in the region where the first order
%transition is suppressed, as shown in the inset of Fig. 2. 
%We find that beyond a
%critical value $g_c'=0.433$, the $N=0$ phase invades the entire phase
%diagram as shown in the Figs.\ \ref{fig1} and \ref{fig2}.  

The phase diagram obtained by numerical solution of the Stoner
criterion is shown in Figs.~1 and 2 for different values of the
dimensionless interaction constant $\tilde g=gN(0)$. Since $\tilde g\propto
1/t_a$, increasing $\tilde g$ can be experimentally achieved by
decreasing pressure \cite{comment2}. For small $\tilde g$, the quantization
of $Q_x$ is essentially exact (see Fig.~\ref{fig1} obtained for
$\tilde g=0.38$). Fig.~\ref{fig2}, which is obtained for $\tilde
g=0.43$, shows that strong deviations from quantization appear for a
sufficiently strong interaction. These deviations are more pronounced
for high values of $N$. For $\tilde g=0.43$, the first-order
transitions $N\simeq 6\to N\simeq 5,N\simeq 7\to N\simeq 6, \cdots$ are
suppressed. The parameter $N=(Q_x-2k_F)/G$ varies continuously in the
corresponding field range. Since $Q_x$ is exactly quantized at $T=0$
(see below), the low-temperature first-order transition line between
the phases $N=6$ and $N=5$ terminates by a second-order critical point
above which the first-order transition is suppressed 
(Fig.~\ref{fig2}). We find that first-order transitions $N+1\to N$ with 
$N<5$ are never suppressed. Indeed, if one increases $\tilde g$
beyond the critical value $\tilde g_c=0.433$, the phase $N=0$ invades
the entire phase diagram. This latter result agrees with the
experimental results showing that the same SDW phase is stable for any
value of the field below a critical pressure $P_c\sim 6$ kbar
\cite{Kang1}.

{\it Zero-temperature phase diagram.} 
To obtain the phase diagram at $T=0$, one should calculate the
condensation energy of the system and look for its
minimum as a function of ${\bf 
Q}$ (at fixed electron density). According to the QNM, each FISDW
phase is characterized by a series of gaps
$\Delta_n=gI_n^2\Delta$ where $\Delta=\langle
\psi^\dagger_{\downarrow-}({\bf r}) \psi_{\uparrow+}({\bf r})\rangle
e^{-i {\bf Q}\cdot {\bf r}}$ is the SDW order parameter. The gap with
the largest amplitude, $\Delta_N$, opens up at the Fermi energy. Here
we allow for a non-quantized wave vector $Q_x=2k_F+NG+z/v_F$ ($N$
integer) and assume that $|z|\ll \Delta_N$. If $Q_x$ is not quantized
($z\neq 0$), the particle number 
conservation implies a shift $\delta\mu=z+{\rm
  sgn}(z)(z^2+\Delta^2_N)^{1/2}$ of the chemical potential. As a result,
the chemical potential does not lie in a gap (since $|\delta\mu|>\Delta_N$),
and the Hall conductance 
is not quantized. In order to determine the value of $z$, we use the
method of Ref.~\cite{Mon1}. We take into account the main gap $\Delta_N$
exactly, and consider the gaps $\Delta_{n\neq N}$ which open away from
the Fermi level within perturbation theory. Skipping technical
details, we obtain the condensation energy 
%%%%%%%%%%%%%%%%%%%%%%%%%%%%%%%%%%%%%%%%%%%%%%%%%%%%%%%%%%%
\begin{figure}
\centerline{\psfig{file=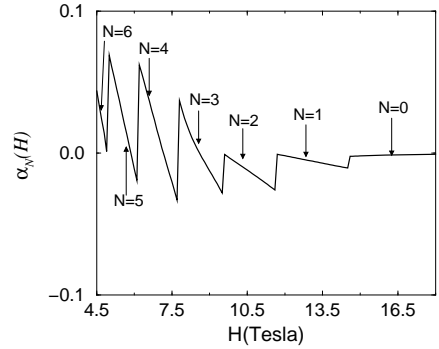,width=0.7\linewidth,angle=0}}
\caption{$\alpha_N(H)$ [Eq.~(\ref{fcond})] {\it vs} field for $\tilde
  g=0.43$. The condition $\alpha_N<1$ 
  implies that quantization of $Q_x$ and hence the Hall conductance is
  exact at $T=0$. }
\label{fig3}
\end{figure}
\noindent
%%%%%%%%%%%%%%%%%%%%%%%%%%%%%%%%%%%%%%%%%%%%%%%%%%%%%%%%%%%
\begin{eqnarray}
\Delta E_{N} &=& -\frac{N(0)}{2}\Delta_N^2
+N(0) \Bigl[|z| (z^2+ \Delta_N^2)^{1/2}-z^2\Bigr] , 
\label{energy1}\\
\frac{2}{\tilde g I_N^2} &=& \ln \frac{2E_0}{\Delta_N}
+\sum_{n\ne0} \frac{I_{N+n}^2}{I_N^2} \ln \frac
{2E_0}{|n\omega_c|} \nonumber\\
&& -{\rm argsh} \frac{|z|}{\Delta_N} - \frac{z}{\omega_c}
\sum_{n\ne 0} \frac{I_{N+n}^2}{nI_N^2} .
\label{deltan}
\end{eqnarray}
Eq.~(\ref{energy1}) shows that for a given value of $\Delta_N$ the
energy is minimum for $z=0$. Therefore, in order to stabilize a phase
with $z\neq 0$, a necessary condition is $\Delta_N(|z|>0) > \Delta_N(z=0)$,
i.e. [see Eq.~(\ref{deltan})]
\begin{eqnarray}
{\rm argsh} \frac{|z|}{\Delta_N} + \sum_{n\ne 0}   \frac {z}{\omega_c}
\frac{I_{N+n}^2}{nI_N^2}  < 0.
\label{cond}
\end{eqnarray}
From Eq.~(\ref{cond}), we conclude that a sufficient condition for
$Q_x$ to be quantized is 
\begin{eqnarray}
\alpha_N(H) &=& \Bigg| \frac {\Delta_N}{\omega_c}
\sum_{n\ne 0} \frac{I_{N+n}^2}{n I_N^2} \Bigg| < 1,
\label{fcond}
\end{eqnarray}
where we have used $|z|\ll \Delta_N$. Given that the $I_n$
coefficients satisfy the sum rule $\sum_n I_n^2=1$ and
$\Delta_N\lesssim \omega_c$ \cite{Mon1}, we expect the
inequality (\ref{fcond}) to be satisfied. Our numerical results for
$\tilde g=0.43$ confirm this 
expectation (Fig.~\ref{fig3}). We find that $\alpha_N(H)$ increases with
$N$, but is always much less than unity. For $\tilde g< 0.43$,
$\alpha_N$ further decreases. We therefore conclude that, while it
is never quantized near $T_c$ for $N\neq 0$ [see Eq.~(\ref{epsilon})], $Q_x$
is strictly quantized at $T=0$ for all values of $N$.

%%%%%%%%%%%%%%%%%%%%%%%%%%%%%%%%%%%%%%%%%%%%%%%%%%%%%%%%%%%
\begin{figure}
\centerline{\psfig{file=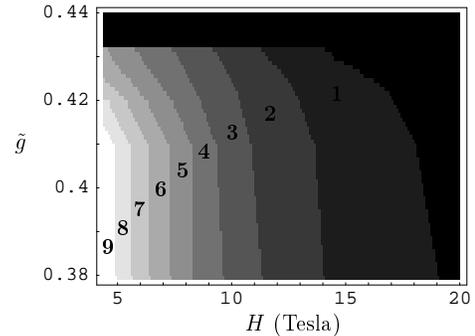,width=0.7\linewidth,angle=0}}
\caption{Zero-temperature phase diagram showing the quantum number $N$ {\it
  vs.} field and interaction strength $\tilde g$. For $\tilde g>\tilde
  g_c=0.433$, the phase 
  $N=0$ (black area) becomes stable for all values of $H$. }
\label{fig4}
\end{figure}
\noindent
%%%%%%%%%%%%%%%%%%%%%%%%%%%%%%%%%%%%%%%%%%%%%%%%%%%%%%%%%%%

The zero-temperature phase diagram, obtained by solving
Eqs.~(\ref{energy1}-\ref{deltan}) is shown in Fig.~\ref{fig4} as a
function of field and electron-electron interaction strength. For
$T=0$, we find that the $N=0$ phase again invades the phase diagram at
$\tilde g_c=0.433\pm 0.001$.  At low temperature, corrections to 
the $T=0$ condensation energy will be exponentially small ($\propto
e^{-\Delta_N/T}$). Thus, the quantization of $Q_x$ will persist in
a finite temperature range.  This implies that for $\tilde g\lesssim
\tilde g_c$, at some intermediate temperature $T^{*}(g,H)$  
between $T=0$ and $T=T_c$, there must be a phase transition or a crossover 
between phases with quantized and non-quantized $Q_x$. This
phase transition/crossover is schematically indicated 
by a dotted line in Fig.~2. The details of this
transition/crossover is beyond the scope of our present study.

{\it Comparison with previous theoretical results.} 
The overall phase diagram that we obtain is therefore qualitatively
different from those obtained in the previous studies
\cite{Gorkov84,Heritier84,Yamaji85,Montambaux85,Virosztek86,Mon1,Lebed}.
Near the 
metal-FISDW transition, we find that $Q_x$ deviates from its
quantized values, in accordance with Ref.~\cite{Lebed}. However, in
contrast to Ref.~\cite{Lebed}, our study indicates that this
deviation is large enough to suppress the first-order transitions only in a
very limited region of the phase diagram corresponding to high values
of $N$ and $\tilde g$ close to $\tilde g_c$ (Fig.~\ref{fig2}). 
In Ref.~\cite{Lebed}, it is assumed that $Q_y=\pi/b$. When $N\neq 0$,
this assumption is not correct and one has to look for the value of
$Q_y$ which maximizes the transition temperature. Even with the
assumption $Q_y=\pi/b$, we are unable to reproduce Lebed's
results. Instead of the FISDW cascade, we find that only the phase
$N=0$ is stable at low temperature albeit with a very low $T_c$. 
Furthermore, at
low temperature ($T\ll T_c$), we find that the quantization of $Q_x$
is exact (implying the quantization of the $T=0$ Hall conductance),
which contradicts the prediction of Ref.~\cite{Lebed} based 
on an extrapolation of results obtained near $T_c$.
This suggests a novel phase transition/crossover at
intermediate temperatures between phases with quantized and
non-quantized $Q_x$. 
Also, below a  critical pressure, we find that the $N=0$ phase invades
the entire phase diagram.

{\it Comparison with experiments.}
The overall phase diagram that we obtain agrees
with the experimental observations in the compound (TMTSF)$_2$PF$_6$
\cite{Kang1}. Above a critical pressure $P_c$ (which corresponds
to $\tilde g<\tilde g_c$ in our theoretical analysis), we describe the
cascade of FISDW phases. When $P<P_c$ we find that the phase $N=0$
invades the entire phase diagram. Thus our study shows that the SDW
phase below $P_c$ is nothing else but the phase $N=0$ of the FISDW cascade
(Fig.~\ref{fig4}). This is also the conclusion obtained in
Ref.~\cite{Kang1}. To our knowledge, the sudden
disappearance of the FISDW cascade below the critical pressure $P_c$ has not
been explained before.

Recent magnetoresistance measurements by Kornilov {\it et al.}
\cite{Korni} found that hysteretic behavior occurs at low temperature
at the transitions 
between successive FISDW phases. The hysteresis weakens at
higher temperature and disappears above a characteristic temperature
$T_0$ ($T_0<T_c$) for all $N>0$. This behavior was ascribed to the
suppression of the first-order transitions in the temperature range
$T_0\lesssim T\lesssim T_c$ in agreement with Lebed's predictions
\cite{Lebed}. However, this interpretation is inconsistent with
our result that the first-order phase transitions can be suppressed
only for $N\geq 5$. We cannot exclude, even if it seems quite unlikely,
that in a more realistic model (for instance taking account of the
triclinic structure of the Bechgaard salts) the suppression of the
first-order phase transitions would also occur for $N<5$. In our
opinion, the conclusion that the absence of hysteresis observed in
experiments originates from the suppression of the first-order
transitions should be taken cautiously. Such an absence of hysteresis
could also be due to the weak first-order character of the transitions
near $T_c$ as was originally thought \cite{Cooper}. Our results
suggest to perform experimental 
studies close to $P_c$, since the suppression of the
first-order transitions should primarily be observed in the close
vicinity of the critical pressure $P_c$ (i.e. $P\gtrsim P_c$) below
which the FISDW cascade disappears.

KS thanks S. Girvin for support under grant DMR-0196503. ND thanks
G. Montambaux and D. J\'erome for useful discussions.

%%%%%%%%%%%%%%%%%%%%%%%%%%%%%%%%%%%%%%%%%%%%%%%%%%%%%%%%%%%%%%%%%%%%

%\vspace{-0.4cm}


\begin{thebibliography}{99}
\bibitem{comment2} An increase in pressure also increases
         $t_{2b}$. This has a similar effect on $T_c$ as a decrease of $\tilde
         g$. Since we do not know how these two parameters
         quantitatively vary with pressure, it is sufficient to vary
         $\tilde g$ (at fixed $t_{2b}$) to mimic a change of pressure.
\end{thebibliography}
\end{document}